\documentclass[aps,prl,twocolumn,showpacs,byrevtex]{revtex4}
\let\oldhat\hat
\renewcommand{\hat}[1]{\oldhat{\mathbf{#1}}}

\usepackage{times,mathptm}
\usepackage[dvips]{graphicx,color}
\usepackage{titlesec}
\newcommand{\bqa}{\begin{eqnarray*}}
\newcommand{\eqa}{\end{eqnarray*}}
\begin{document}
\title{Hidden Weyl Fermions in Paramagnetic Electride Y$_2$C}
\author{Liangliang Liu$^{1,2}$, Chongze Wang$^1$, Seho Yi$^1$, Dou Kyun Kim$^3$, Chul Hong Park$^3$, and Jun-Hyung Cho$^{1*}$}
\affiliation{$^1$Department of Physics, Research Institute for Natural Science, and HYU-HPSTAR-CIS High Pressure Research Center, Hanyang University, Seoul 133-791, Korea \\
$^2$ Key Laboratory for Special Functional Materials of Ministry of Education, Henan University, Kaifeng 475004, People's Republic of China \\
$^3$ Department of Physics Education, Pusan National University, Pusan 609-735, Korea}
\date{\today}

\begin{abstract}
Recent experimental observations of Weyl fermions in materials opens a new frontier of condensed matter physics. Based on first-principles calculations, we here discover Weyl fermions in a two-dimensional layered electride material Y$_2$C. We find that the Y 4$d$ orbitals and the anionic $s$-like orbital confined in the interstitial spaces between [Y$_2$C]$^{2+}$ cationic layers are hybridized to give rise to van Have singularities near the Fermi energy $E_{\rm F}$, which induce a ferromagnetic (FM) order via the Stoner-type instability. This FM phase with broken time-reversal symmetry hosts the rotation-symmetry protected Weyl nodal lines near $E_{\rm F}$, which are converted into the multiple pairs of Weyl nodes by including spin-orbit coupling (SOC). However, we reveal that, due to its small SOC effects, Y$_2$C has a topologically nontrivial drumhead-like surface state near $E_{\rm F}$ as well as a very small magnetic anisotropy energy with several ${\mu}$eV per unit cell, consistent with the observed surface state and paramagnetism at low temperatures below ${\sim}$2 K. Our findings propose that the Brillouin zone coordinates of Weyl fermions hidden in paramagnetic electride materials would fluctuate in momentum space with random orientations of the magnetization direction.
\end{abstract}
\maketitle


As an emerging class of low-dimensional electron systems, electrides have attracted considerable attention because of their promising prospects in both fundamental research and technological applications~\cite{ele1,ele2,ele3,ele4}. In electrides, the loosely-bound electrons are easily separated from cationic atoms, thereby being trapped in void spaces along one-dimensional channels~\cite{ele5,ele6} or between two-dimensional (2D) interlayers~\cite{ele7,seho}. Such low-dimensional anionic electrons occupying the bands near the Fermi level $E_{\rm F}$ may provide unconventional playgrounds for exploration of various exotic quantum phenomena such as charge-density waves, spin ordering, superconductivity, and topological states~\cite{Gruner,spin,supercond,topo2}. Recently, Lee $et$ $al$. demonstrated the synthesis of a layered electride material Ca$_2$N, where the anionic electrons are distributed in the interlayer spaces between positively charged [Ca$_2$N]$^{+}$ cationic layers~\cite{Lee}. After such a pioneering realization of 2D electride, extensive searches have been theoretically and experimentally carried out to find various types of 2D electride materials that offer the unique properties of high electrical conductivities~\cite{elecon}, low work functions~\cite{Ming}, highly anisotropic optical response~\cite{Lee}, and efficient catalysts~\cite{catalysis}.

\begin{figure}[h!t]
\includegraphics[width=8cm]{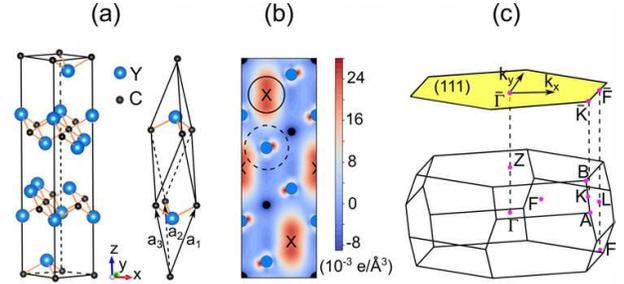}
\caption{(Color online) (a) Optimized structure of bulk Y$_2$C within the trigonal conventional cell (left) and rhombohedral primitive cell (right), (b) spin density plotted on the (110) face of the conventional cell, and (c) Brillouin zone (BZ) of the rhombohedral primitive cell together with the projected surface BZ of the (111) surface. In (b), ``X" denotes the central region of localized anionic electrons, and the dashed (solid) circle represents the sphere of radius 1.82 (1.50) {\AA} around the Y (X) atom. }
\label{figure:1}
\end{figure}

Among several existing 2D layered electrides, Y$_2$C containing two Y atoms and one C atom within the rhombohedral primitive unit cell [see Fig. 1(a)] shows a semimetallic feature with the electron and hole pockets near $E_{\rm F}$~\cite{hole}. Despite the recent intensive studies of Y$_2$C~\cite{hole,low,Otani,pressure,Hiraishi,high,electride,y2cnl,Ferro,Ferro1}, there still remains a strong discrepancy for its ground state between the experimental measurements~\cite{hole,low,Otani,pressure,Hiraishi,Ferro1} and theoretical predictions~\cite{high,electride,y2cnl,Ferro}. According to experiments~\cite{low,Otani}, Y$_2$C exhibits paramagnetism even at low temperatures below ${\sim}$2 K. However, this observed paramagnetism has not so far been properly explained by previous density-functional theory (DFT) calculations which either assumed a nonmagnetic (NM) ground state~\cite{high,electride,y2cnl} or predicted a ferromagnetic (FM) ground state with large magnetic anisotropy~\cite{Ferro,Ferro1}. Nevertheless, all theories~\cite{high,electride,y2cnl,Ferro,Ferro1} agree that there are highly localized charges or spins in the interstitial spaces between [Y$_2$C]$^{2+}$ cationic layers [see Fig. 1(b)].

In this Letter, based on first-principles calculations~\cite{method}, we report that Y$_2$C behaves as paramagnetic with nearly degenerate FM Weyl semimetal states. By analyzing the electronic structures of few-layer and bulk Y$_2$C, we find that the FM phase begins to be stabilized from bilayer Y$_2$C via the Stone-type instability, indicating that the confined anionic electrons between the two [Y$_2$C]$^{2+}$ cationic layers are associated with the appearance of FM spin ordering. For bulk Y$_2$C, the hybridization of Y 4$d$ orbitals and anionic $s$-like orbital produces small orbital angular momenta, which in turn contribute to a magnetic anisotropy energy (MAE) of several ${\mu}$eV per unit cell. This extremely small MAE invokes FM fluctuations which can provide an explanation for the experimental observation~\cite{low,Otani,Hiraishi} of paramagnetism even at ${\sim}$2 K. Remarkably, the electronic structure of the FM phase shows the existence of the rotation-symmetry protected Weyl nodal lines near $E_{\rm F}$, which are converted into the multiple pairs of Weyl nodes by including spin-orbit coupling (SOC). In particular, we identify a drumhead-like surface state near $E_{\rm F}$, the dispersion of which is insensitive to the positions of Weyl nodes varying with respect to the magnetization direction. This topologically nontrivial surface state is corroborated by a previous angle-resolved photoemission spectroscopy (ARPES) measurement~\cite{hole}. Thus, our findings not only solve the outstanding discrepancy between experiment and theory regarding the ground state of Y$_2$C, but also illustrate the exploration of Weyl fermions whose Brillouin zone coordinates fluctuate in momentum space.

We begin by examining the relative stability of the FM and NM phases in few-layer Y$_2$C with increasing the number $N$ of layers. Figure 2(a) shows the calculated energy difference ${\Delta}E_{\rm FM-NM}$ between the FM and NM phases as a function of $N$. We find that a monolayer (ML) Y$_2$C ($N$ = 1) has the NM ground state, while few-layer Y$_2$C with $N$ ${\ge}$ 2 have the FM one. It is noted that for $N$ = 2, the FM ground state is more stable than the antiferromagnetic state by ${\sim}$0.5 meV per ML (see Fig. S1 of the Supplemental Material~\cite{SM}). In Fig. 2(a), the calculated magnetic moment $m$ of the FM phase is also displayed with respect to $N$. It is seen that $m$ increases monotonously with increasing $N$, being saturated to be 0.383 ${\mu}_{\rm B}$/ML at bulk Y$_2$C. To explore the underlying mechanism of the FM order, we calculate the band structure and the density of states (DOS) for the NM phase of bilayer ($N$ = 2) and bulk Y$_2$C. For $N$ = 2, the calculated band structures exhibit the electron (hole) pocket around the $\overline{\rm F}$ ($\overline{\rm K}$) point near $E_{\rm F}$ [see Fig. 2(b)], producing the van Have singularities (vHs) with a large total DOS [see the inset of Fig. 2(c)]. Consequently, the FM order is induced via Stoner criterion $D(E_{\rm{F}})I > 1$~\cite{stoner} (see Fig. S2 of the Supplemental Material~\cite{SM}), where $D(E_{\rm{F}})$ is the total DOS at $E_{\rm F}$ and the Stoner parameter $I$ can be estimated with dividing the exchange splitting of spin-up and spin-down bands by the corresponding magnetic moment.

\begin{figure}[h!t]
\includegraphics[width=8cm]{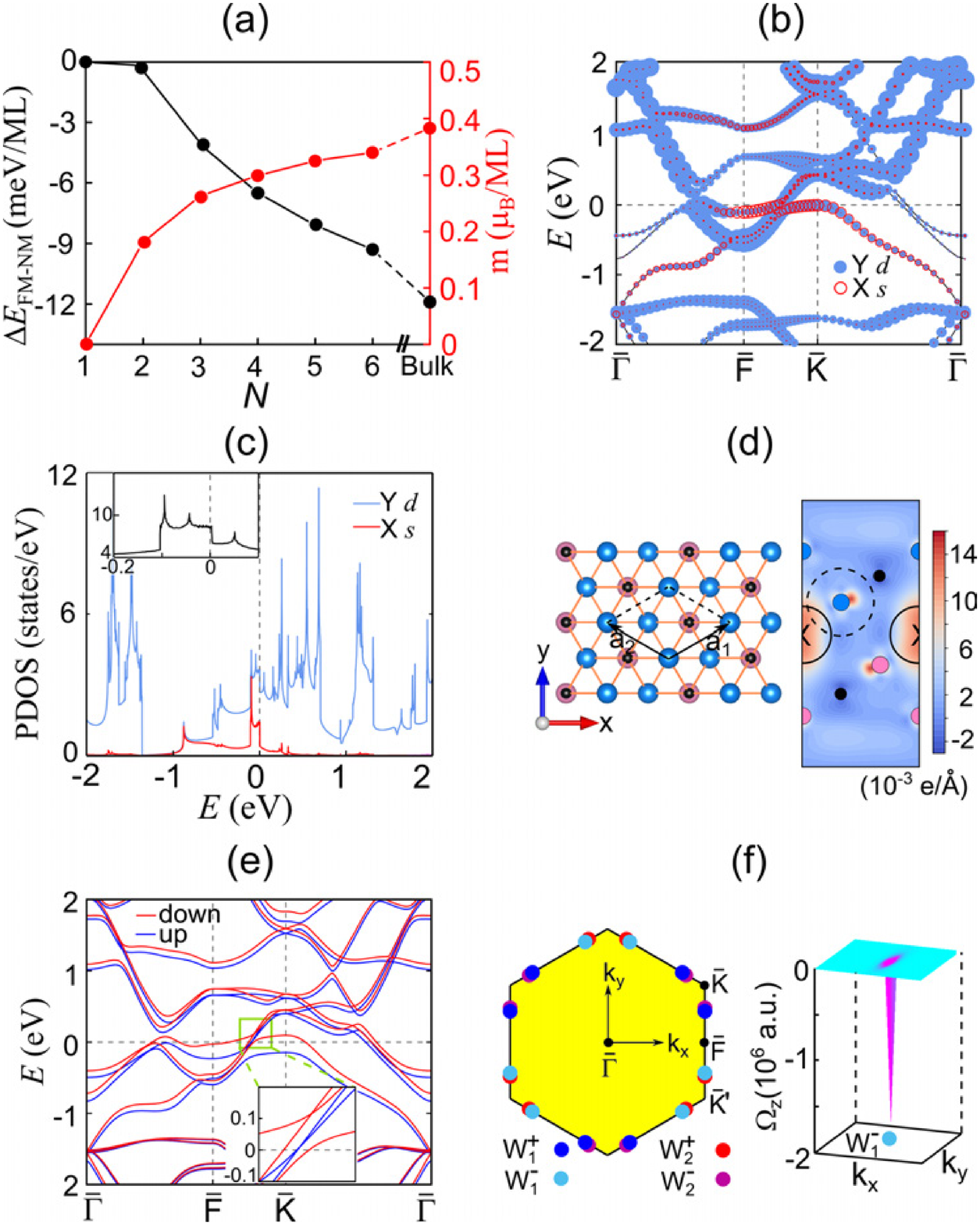}
\caption{(Color online) (a) Calculated energy difference ${\Delta}E_{\rm FM-NM}$ and magnetic moments $m$ of the FM phase as a function of $N$.
For bilayer Y$_2$C, the calculated band structure and PDOS of the NM phase are displayed in (b) and (c), respectively. In (b), the bands projected onto the Y $d$ and X $s$ orbitals are displayed with circles whose radii are proportional to the weights of the corresponding orbitals. The energy zero represents $E_{\rm F}$. A close-up of the total DOS near $E_{\rm F}$ is given in the inset of (c). The structure (top view) and spin density of bilayer Y$_2$C is drawn in (d), where the blue and pink circles represent the first- and second-layer Y atoms. The band structure for the FM phase of bilayer Y$_2$C is given in (e), together with the zoom-in band structure along the $\overline{\rm FK}$ line. In (f), three nonequivalent pairs of Weyl nodes for each spin channel are drawn in the BZ and the  distribution of the Berry curvature component $\Omega_{z}$ around W$_{1}^{-}$ is also displayed.
}
\label{figure:2}
\end{figure}

It is noteworthy that for bilayer Y$_2$C, the partial DOS (PDOS) projected onto the Y 4$d$ orbitals and the anionic $s$-like orbital confined in the interstitial regions between the two [Y$_2$C]$^{2+}$ cationic layers exhibits sharp peaks close to $E_{\rm F}$ [see Fig. 2(c)], indicating a strong hybridization of the two orbitals. Compared to other orbitals, these two orbitals are found to be more dominant components of the electron- and hole-pocket states near $E_{\rm F}$: see the band projections in Fig. S3 of the Supplemental Material~\cite{SM}. Due to such a hybridization of the Y 4$d$ and anionic $s$-like orbitals, the spin densities for bilayer and bulk Y$_2$C are distributed over the Y atoms and the interstitial regions X, as shown in Figs. 2(d) and 1(b), respectively. For bilayer (bulk) Y$_2$C, the calculated spin moments integrated within the spheres around Y and X [see Figs. 2(d) and 1(b)] are 0.043 (0.108) and 0.088 (0.169) ${\mu}_{\rm B}$, respectively. Based on our results, we can say that for $N$ ${\geq}$ 2 and bulk Y$_2$C, the vHs arising from the electron- and hole-pockets with the hybridization of the Y 4$d$ and anionic $s$-like orbitals cause the Stoner-type instability to induce an FM spin ordering.

Figure 2(e) shows the band structure for the FM phase of bilayer Y$_2$C. We find that there exist two spinful Weyl nodes just above $E_{\rm F}$ along the $\overline{\rm FK}$ line [see the inset of Fig. 2(e)]. It is noted that the crystalline symmetries of bilayer Y$_2$C represent the point group $D_{3d}$, which contains inversion symmetry $I$, threefold rotational symmetry $C_{3z}$ about the $z$ axis, and twofold rotation symmetry $C_{2y}$ about the $y$ axis. Therefore, for each spin channel, we have three nonequivalent pairs of Weyl nodes along the $\overline{\rm FK}$ and $\overline{\rm FK'}$ lines [see Fig. 2(f)], where each Weyl node at a point ${\bf k}$ is paired with the other Weyl node of opposite chirality at $-$${\bf k}$~\cite{weyl1,weyl2}. Note that the twofold degeneracy of such 2D Weyl nodes (located along the $k_y$ direction) is protected by $C_{2y}$ rotation~\cite{weyl2}: i.e., two crossing bands have opposite eigenvalues ${\pm}1$ of $C_{2y}$. In order to verify these Weyl nodes, we calculate the Berry curvature around the band touching points by using the WannierTools package~\cite{wanniertool}. Here, the Wannier bands are in good agreement with the first-principles bands (see Fig. S4 of the Supplemental Material~\cite{SM}). It is found that each pair of Weyl nodes have the positive and negative Berry curvature distributions [see Fig. 2(f)], which can be regarded as the source and sink of Berry curvature in momentum space, respectively.

It is interesting to examine how the Weyl nodes in bilayer Y$_2$C evolve as such 2D Weyl semimetal is stacked into bulk Y$_2$C. Figure 3(a) shows the band structure for the FM phase of bulk Y$_2$C. Similar to bilayer Y$_2$C, there are two spinful Weyl nodes along the $\overline{\rm LA}$ line (parallel to $\overline{\rm FK}$ in 2D BZ), which in turn form closed nodal loops around the L points [see Fig. 3(b)]. The presence of such nodal loops with large deformation along the $k_z$ direction implies strong interlayer couplings between cationic layers through anionic electrons, contrasting with small deformed nodal lines reported in layered materials with weak van der Waals interlayer couplings~\cite{Nie}. Since the crystalline symmetries of bulk Y$_2$C belong to the space group $R\overline{3}m$ (No. 225) with the point group $D_{3d}$, there are three separate Weyl nodal lines (WNLs) for each spin channel. Here, the twofold degeneracy of WNL (located along the $k_x$ direction) is protected by $C_{2y}$ rotation, similar to the case of Weyl nodes in bilayer Y$_2$C. To confirm such symmetry protection of WNLs, we introduce various perturbations of the Y atoms which break or preserve three nonequivalent $C_2$ rotation symmetries. Our calculated band structures show a gap opening of the WNLs (see Fig. S5 of the Supplemental Material~\cite{SM}) only for the symmetry-broken geometries, confirming that the gapless WNLs are protected by the $C_2$ symmetries. We further demonstrate the topological characterization of the WNLs by calculating the topological index~\cite{Z2index}, defined as ${\zeta}_1$ = ${\frac{1}{\pi}}$ ${\oint}$$_c$ $dk$${\cdot}$A($k$), along a closed loop encircling any of the WNLs. Here, A(k) = $-i$$<$$u_k$$\mid$$\partial$$_k$$\mid$$u_k$$>$ is the Berry connection of the related Bloch bands. We obtain ${\zeta}_1$ = ${\pm}$1 for the WNLs, indicating that they are stable against symmetry conserving perturbations.

\begin{figure}[h!t]
\includegraphics[width=8cm]{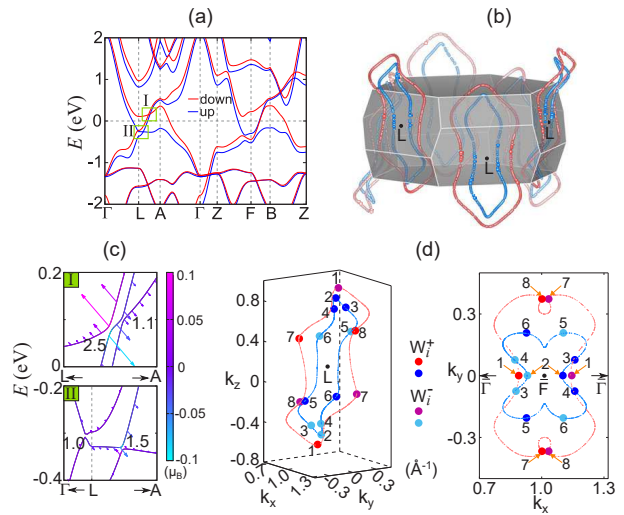}
\caption{(Color online) (a) Band structure and (b) WNLs of the FM phase of bulk Y$_2$C, computed in the absence of SOC. The close-up band structures around the regions I and II, computed with including SOC, are given in (c). Here, the arrows along the horizontal and vertical directions represent the $m_{x}$ and $m_{y}$ components, respectively, while the arrow colors indicate $m_{z}$. The values $m_{x}$, $m_{y}$, and $m_{z}$ for the longest arrow are $-$0.117, 0.067, and 0.109 ${\mu}_{\rm B}$, respectively. In (c), the numbers represent the SOC-induced gap (in meV) at several ${\bf k}$ points. In (d), the distribution of Weyl nodes is drawn in the 3D (left) and 2D (right) views.
}
\label{figure:3}
\end{figure}

So far, we have considered the band structures in the absence of spin-orbit coupling (SOC), where two spin channels in the FM phase are decoupled from each other because of the independence of the spin and orbital degrees of freedom. However, the inclusion of SOC lifts the degeneracy at the band-crossing points along the $\overline{\rm LA}$ line, as shown in Fig. 3(c). Each spin-up (spin-down) WNL in Fig. 3(b) becomes gapped with the exception of three (five) pairs of Weyl nodes [see Fig. 3(d)]. The positions of Weyl nodes in momentum space are given in Table SI of the Supplemental Material~\cite{SM}, together with their energies. Here, the spontaneous magnetization direction is calculated to be along the $z$ axis, consistent with the experimental measurement of anisotropic magnetic properties~\cite{Ferro1}. The corresponding magnetic point group is C$_{3i}$ containing $I$, $C_{3z}$ about the $z$ axis, as well as the product $C_{2y}T$ of rotation $C_{2y}$ and time reversal $T$. Therefore, the WNLs protected by three nonequivalent $C_{2}$ symmetries are not allowed anymore, leading to the opening of SOC gaps depending on unquenched orbital magnetic moments [see Fig. 3(c)]. As shown in Fig. 3(d), each pair of Weyl nodes related by inversion $I$ has its counterpart through $C_{2y}T$ symmetry: e.g., (W$_{3}^{+}$, W$_{3}^{-}$) and (W$_{4}^{+}$, W$_{4}^{-}$). Note that such counterparts of (W$_{1}^{+}$, W$_{1}^{-}$) and (W$_{2}^{+}$, W$_{2}^{-}$) are themselves.

Figure 4(a) plots the FM band structure along the M$-$L$-$M$_1$ line involving the W$_{1}^{+}$ and W$_{1}^{-}$ nodes of positive and negative chiralities, respectively. The 2D views of Berry curvature around W$_{1}^{+}$ and W$_{1}^{-}$ are displayed in Fig. 4(b). We determine the chirality of each Weyl node by integrating the Berry curvature through a closed 2D manifold enclosing the node. The computed chirality (i.e., the Chern number) is $C$ = +1 and $-$1 for W$_{1}^{+}$ and W$_{1}^{-}$, respectively. Since the hallmark of Weyl nodes is the existence of topologically protected surface states, we calculate the surface electronic structure of Y$_2$C using the Green's function method based on the tight-binding Hamiltonian with maximally localized Wannier functions~\cite{wanniertool,wannier90}. Figure 4(c) shows the projected surface spectrum on the (111) surface of Y$_2$C. Obviously, we find a topological surface state connecting two Weyl nodes W$_{1}^{+}$ and W$_{1}^{-}$ around the $\overline{F}$ point. In Fig. 4(e), we plot the projected Fermi surface of the (111) surface, obtained at a chemical potential of $-$0.183 eV below $E_{\rm F}$. A close-up of this Fermi surface represents two Fermi arcs connecting the W$_{1}^{+}$ and W$_{1}^{-}$ nodes, showing the same shape as the drumhead surface state in the WNL semimetal state obtained without SOC [see Fig. S6(a) of the Supplemental Material~\cite{SM}]. Note that this drumhead shape remains invariant as the chemical potential is lowered up to $-$0.215 eV [see Fig. S6(b)]. It is thus likely that the small SOC-induced gap openings [see Fig. 3(c)] along the WNLs except at the Weyl nodes hardly change the dispersion of the drumhead surface state~\cite{weyl2}.

\begin{figure}[h!t]
\includegraphics[width=8cm]{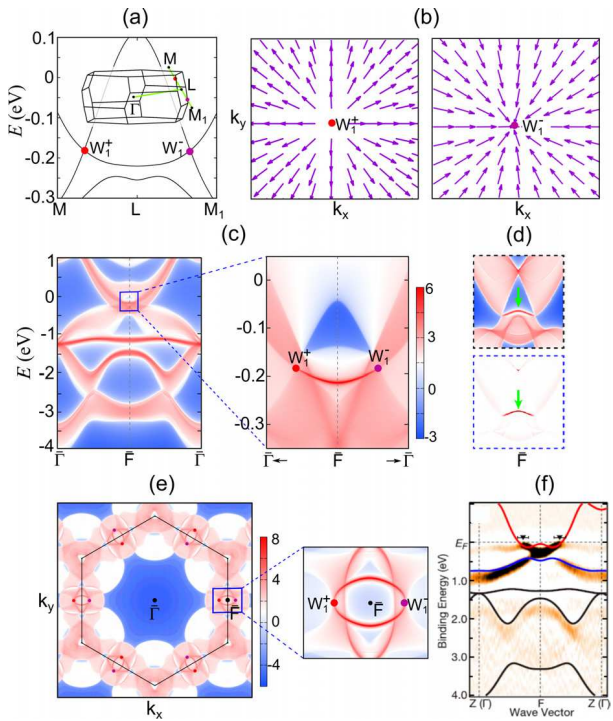}
\caption{(Color online) (a) FM band structure of bulk Y$_2$C along the M$-$L$-$M$_1$ line, (b) Berry curvature around W$_{1}^{+}$ and W$_{1}^{-}$, and
(c) projected surface spectrum for the (111) surface of Y$_2$C with a close-up image around the $\overline{\rm F}$ point. For comparison, the previous theoretical result~\cite{y2cnl} for the projected (111) surface spectrum, obtained from the NM phase, is drawn in (d) (Copyright \copyright 2018 American Chemical Society). The isoenergy surface at $-$0.183 eV below $E_{\rm F}$ is displayed in (e). Experimental band structure~\cite{hole} obtained by plotting the second derivative of the ARPES spectra (Copyright \copyright 2017 American Physical Society) is given in (f).
}
\label{figure:3}
\end{figure}

In Fig. 4(f), the ARPES data~\cite{hole} exhibits a strong intensity around the $\overline{\rm F}$ point between the partially occupied and fully occupied bands near $E_{\rm F}$. Recently, Huang $et$ $al$.~\cite{y2cnl} interpreted such observed in-gap states in terms of a topological surface state [see Fig. 4(d)] originating from the topological property of $Z_2$ invariant in bulk Y$_2$C, which was derived from the assumption of the NM ground state. However, as shown in Fig. 4(d), the dispersion of the surface state is concave downward around the $\overline{\rm F}$ point, contrasting with the measured concave-upward dispersion in the ARPES data [see Fig. 4(f)]. It is remarkable that the present FM topological surface state connecting W$_{1}^{+}$ and W$_{1}^{-}$ shows a concave-upward dispersion [see Fig. 4(c)], in good agreement with the ARPES result obtained from the paramagnetic phase. This implies that the dispersion of the surface state would be insensitive to the magnetization directions, as demonstrated below.

In order to explain why experiments have not observed ferromagnetism even at low temperatures below ${\sim}$2 K~\cite{low,Otani,Hiraishi}, we calculate the MAE for bulk Y$_2$C. We find that the magnetic configuration with the magnetization direction along the $z$ axis is more favorable than that along the $x$ axis by ${\sim}$8 ${\mu}$eV per unit cell. It is noted that other magnetic configurations with different magnetization directions parallel to the $x$-$y$ plane are nearly degenerate. This extremely small MAE reflects a very weak SOC in bulk Y$_2$C, consistent with the experimental observation of no obvious magnetic anisotropy~\cite{Otani,Hiraishi}. Indeed, the magnitudes of orbital magnetic moments are calculated to be two orders smaller than those of spin moments (see Table SII of the Supplemetal Material~\cite{SM}). The resulting tiny magnetic anisotropy provides an explanation for the experimentally observed paramagnetism at low temperatures below ${\sim}$2 K~\cite{low,Otani,Hiraishi}. It is noteworthy that, although the positions of Weyl nodes change depending on the magnetization direction, the drumhead-like surface state around the $\overline{\rm F}$ point remains intact (see Fig. S7 of the Supplemental Material~\cite{SM}). Thus, we can say that the dispersion of the paramagnetic surface state [see Fig. 4(f)], measured by a previous ARPES experiment~\cite{hole}, nearly coincides with that of the topologically nontrivial drumhead surface state generated from the WNL semimetal state.

To conclude, based on first-principles calculations, we have predicted new FM Weyl semimetal states in a 2D layered electride material Y$_2$C. By a systematic study of the electronic structures of few-layer and bulk Y$_2$C, we identified that the hybridization of the Y $d$ orbitals and the anionic $s$-like orbital confined between the two [Y$_2$C]$^{2+}$ cationic layers comprises the vHs near $E_{\rm F}$, therefore inducing an FM spin ordering to produce a Weyl semimetal. In particular, it is revealed that, due to its small SOC effects, Y$_2$C has not only a drumhead-like surface state near $E_{\rm F}$ characterizing WNL semimetal but also a very weak magnetic anisotropy to invoke FM fluctuations, providing an explanation for the observed surface state and paramagnetism at ${\sim}$2 K. The present exploration of Weyl fermions hidden in an apparent paramagnetic electride Y$_2$C manifests the intriguing combination of topology and electride materials. By applying an external magnetic field which can easily switch the magnetization direction of Y$_2$C, it is possible not only to realize the emergence of FM Weyl semimetal states but also to tune the positions of Weyl nodes. Surprisingly, since the lanthanide carbides such as Gd$_2$C, Tb$_2$C, Dy$_2$C, Ho$_2$C, and Er$_2$C~\cite{high} have shown the same/similar FM order, crystalline symmetries, and semimetal band dispersions as those of Y$_2$C, we anticipate that the emergence of FM Weyl semimetal states can be generic to all these lanthanides.

\vspace{0.4cm}

\noindent {\bf Acknowledgements.}
This work was supported by the National Research Foundation of Korea (NRF) grant funded by the Korean Government (Grant Nos. 2016K1A4A3914691 and 2015M3D1A1070609). The calculations were performed by the KISTI Supercomputing Center through the Strategic Support Program (Program No. KSC-2017-C3-0080) for the supercomputing application research and by the High Performance Computational Center of Henan University.  \\

L. L., C. W., and S. Y. contributed equally to this work.


\noindent $^{*}$ Corresponding author: chojh@hanyang.ac.kr

\newpage
\onecolumngrid
\titleformat*{\section}{\LARGE\bfseries}

\renewcommand{\thefigure}{S\arabic{figure}}
\setcounter{figure}{0}

\vspace{1.2cm}

\section{Supplemental Material for "Hidden Weyl Fermions in Paramagnetic Electride Y$_2$C"}
\begin{flushleft}
{\bf 1. Magnetic structures of bilayer Y$_2$C.}
\begin{figure}[ht]
\includegraphics[width=8cm]{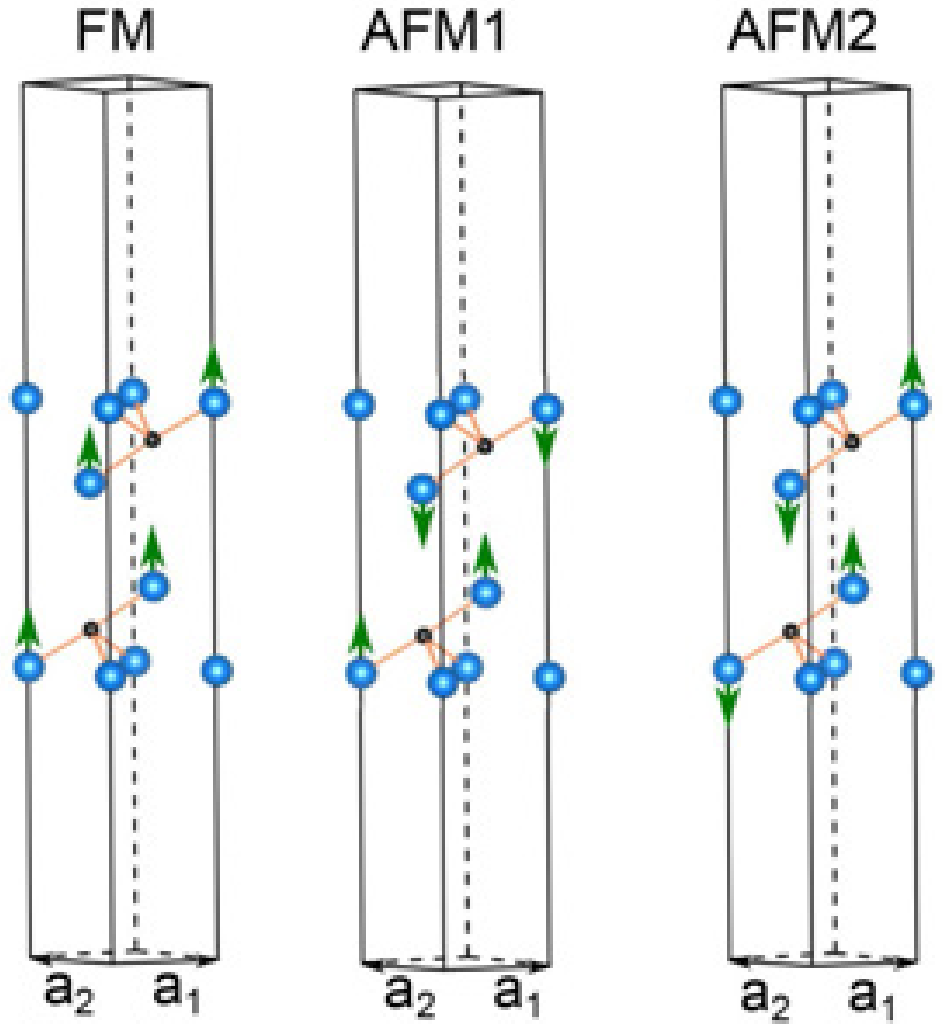}
\caption{Ferromagnetic (FM) and antiferromagnetic (AFM) configurations of bilayer Y$_2$C. Two different AFM configu-rations are considered. Here, the arrows indicate up and down spins. The FM configuration is more stable than the NM one by 0.23 eV per monolayer (ML), while AFM is less stable than NM by 0.27 eV per ML.}
\end{figure}

\vspace{0.4cm}

{\bf 2. Stoner criterion for bilayer and bulk Y$_2$C.}
\begin{figure}[ht]
\includegraphics[width=12cm]{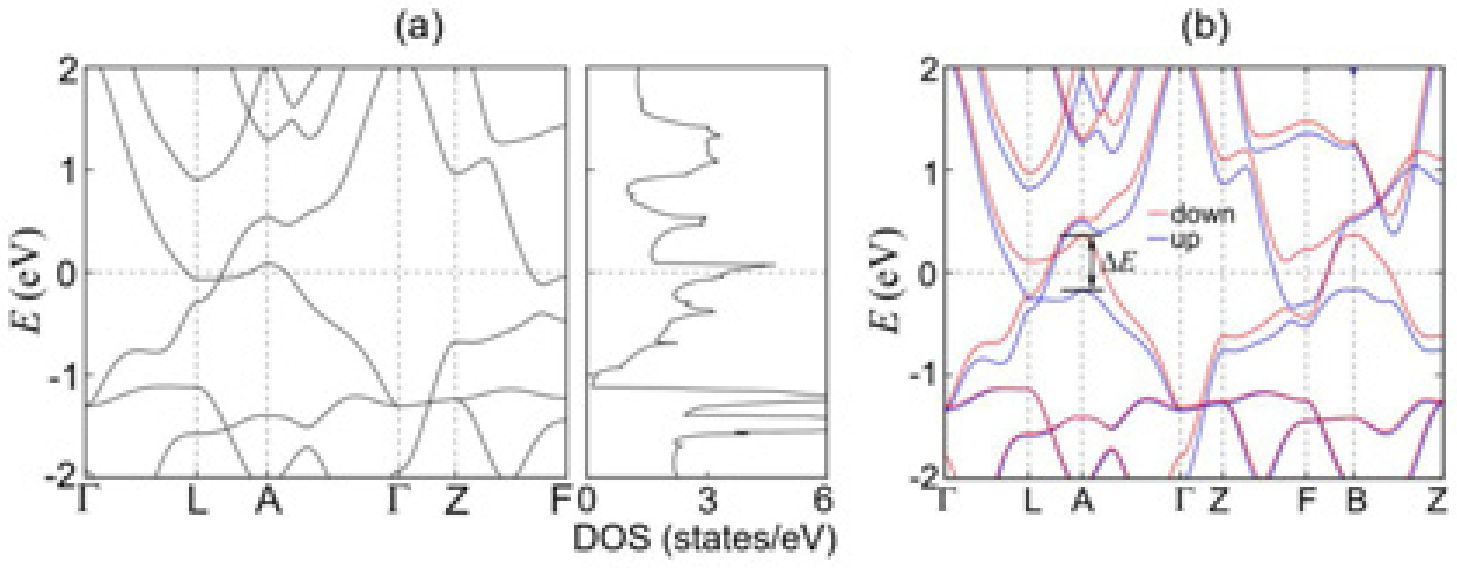}
\caption{(a) Calculated band structure and density of states (DOS) of the NM phase of bulk Y$_2$C. The band structure of the FM phase of bulk Y$_2$C is given in (b), where the exchange splitting (${\Delta}$$E$) of spin-up and spin-down bands are denoted. For bilayer (bulk) Y$_2$C, we obtain $D$($E_{\rm F}$) = 1.76 (1.64) states/eV per spin and $I$ = 1.03 (1.39), giving rise to $D$($E_{\rm F}$)${\cdot}$$I$ = 1.81 (2.27). Therefore, bilayer and bulk Y$_2$C satisfy the Stoner criterion $D$($E_{\rm F}$)${\cdot}$$I$ > 1. Here, the Stoner parameter $I$ for bilayer (bulk) Y$_2$C is estimated with dividing ${\Delta}$$E$ = 0.19 (0.53) eV by the total magnetic moment 0.18 (0.38) ${\mu}$$_{\rm B}$.}
\end{figure}

\newpage

{\bf 3. Band projections onto the Y, C, and X orbitals.}
\begin{figure}[hb]
\includegraphics[width=10cm]{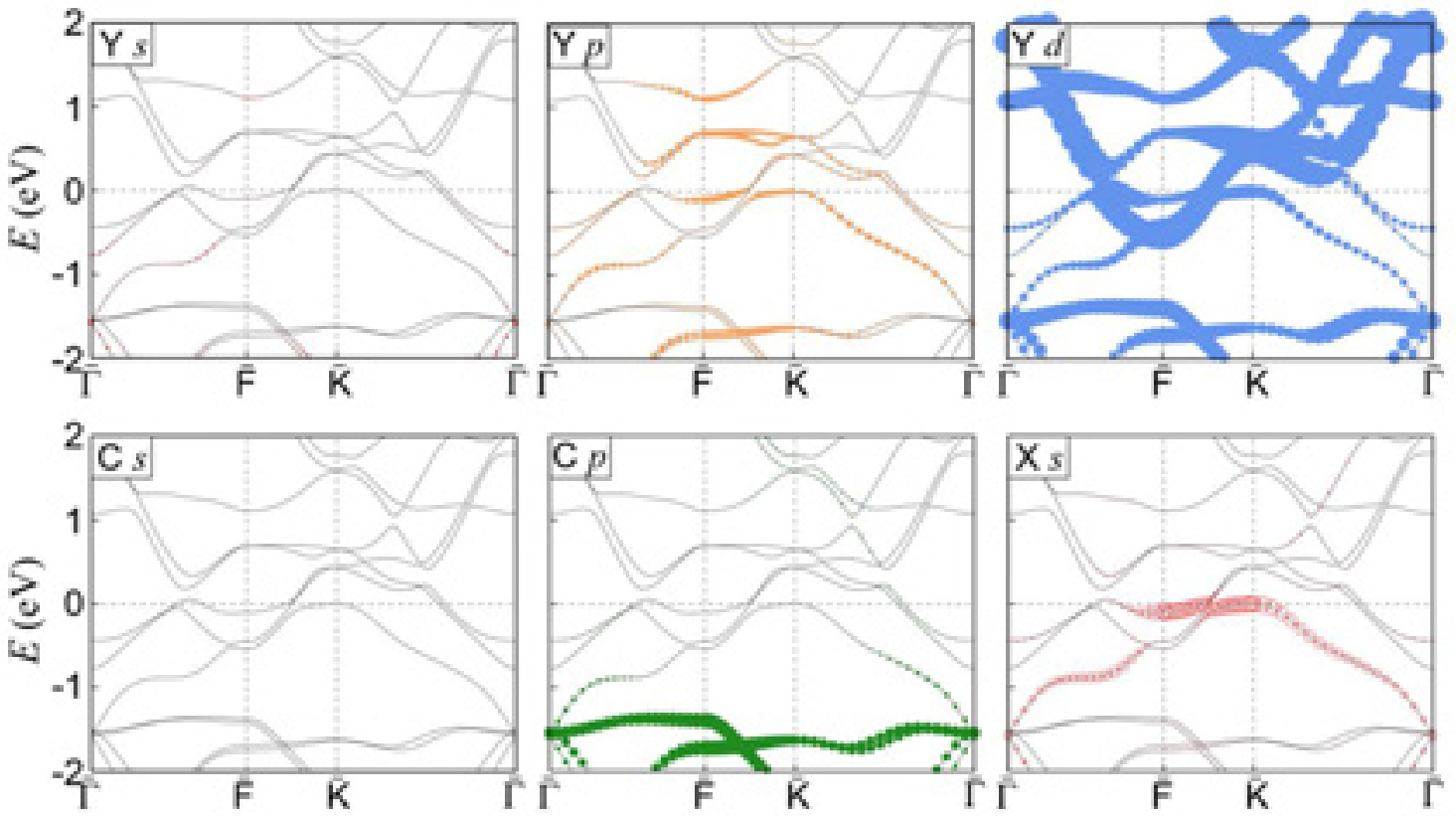}
\caption{Calculated bilayer Y$_2$C bands projected onto the Y $s$, Y $p$, Y $d$, C $s$, C $p$, and X $s$ orbitals. Here, the radii of circles are proportional to the weights of the corresponding orbitals. The Y $d$ and X $s$ orbitals are found to be more dominant components of the electron- and hole-pocket states near the Fermi energy, compared to other orbitals.}
\end{figure}

\vspace{0.4cm}
{\bf 4. Comparison of the electronic bands obtained using the DFT and tight-binding Hamiltonian calculations.}
\begin{figure}[ht]
\includegraphics[width=10cm]{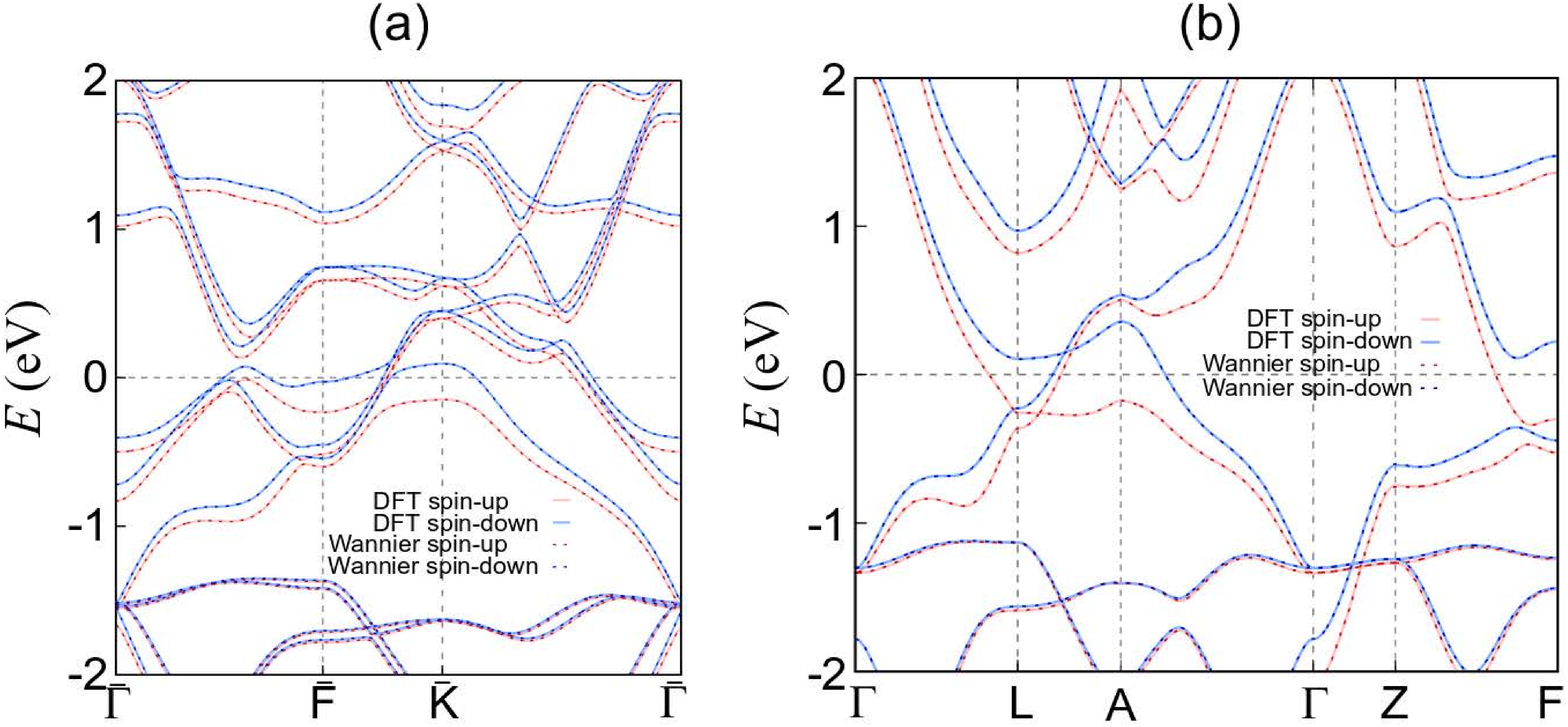}
\caption{Band structures of (a) bilayer and (b) bulk Y$_2$C obtained using the tight-binding Hamiltonian with maximally localized Wannier functions (MLWF), in comparison with that obtained using the DFT calculation. The DFT band structure is well reproduced by using the Wannier90 package~\cite{Wannier90}. Here, the wave functions are projected onto the Y $s$, Y $d$, and C $p$ orbitals.}
\end{figure}

{\bf 5. Band structure of broken C$_2$ rotation symmetry.}
\begin{figure}[ht]
\includegraphics[width=10cm]{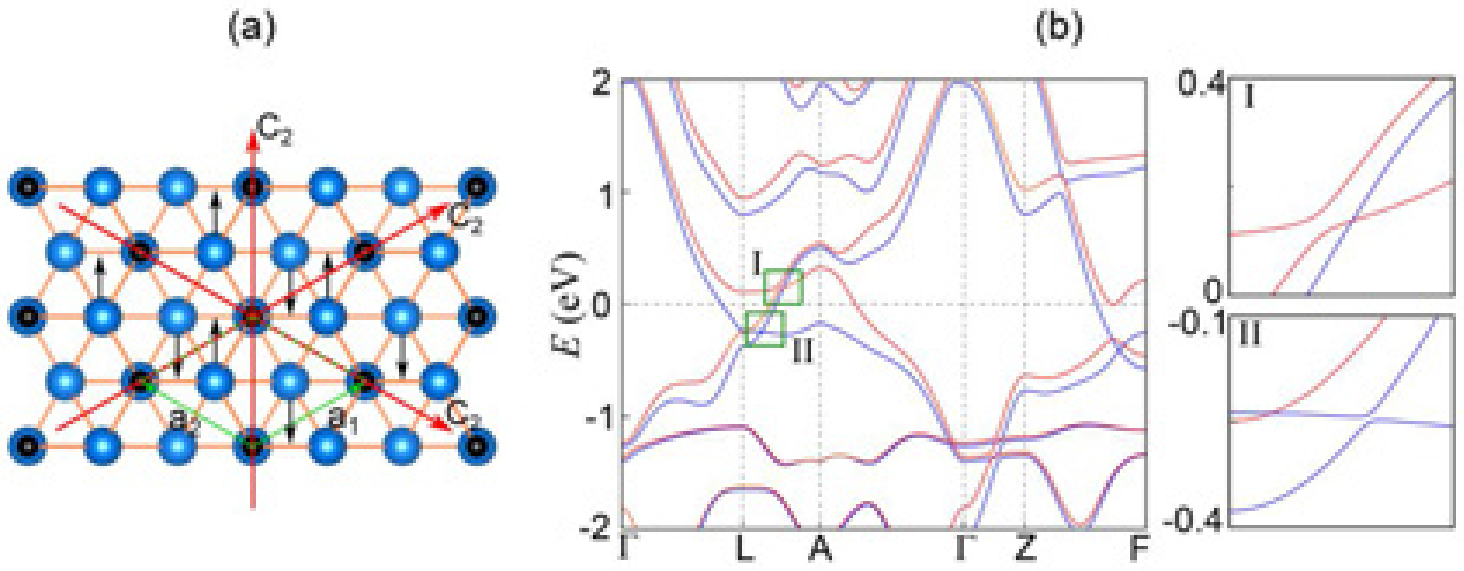}
\caption{(a) Breaking of the three nonequivalent C$_2$ rotation symmetries in bulk Y$_2$C, where Y atoms shift by 0.1 {\AA} (along the arrow directions) from the optimized structure. The calculated band structure for such a symmetry-broken geometry is displayed in (b), showing a gap opening of the WNLs.}
\end{figure}

\newpage

{\bf 6. Drumhead-like surface state.}
\begin{figure}[ht]
\includegraphics[width=14cm]{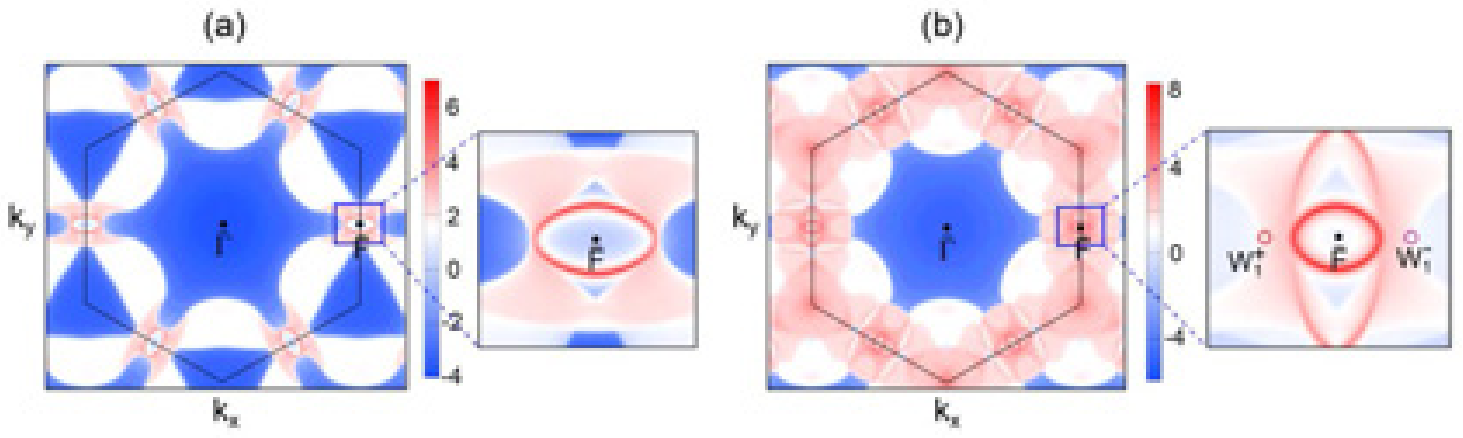}
\caption{(a) Projected Fermi surface of the WNL semimetal state obtained without SOC, obtained at a chemical potential of −0.183 eV below $E_{\rm F}$. The zoom-in isoenergy surface clearly shows the shape of the drumhead-like surface state. In the presence of SOC, the corresponding isoenergy surface at −0.20 eV below $E_{\rm F}$ is displayed in (b), where the drumhead shape remains invariant. The two open circles in (b) represent the projected positions of two Weyl nodes,W$_{1}^{+}$ and W$_{1}^{-}$}
\end{figure}

\vspace{0.4cm}
{\bf 7. Weyl nodes and drumhead surface state, obtained with changing the magnetization direction.}
\begin{figure}[ht]
\includegraphics[width=14cm]{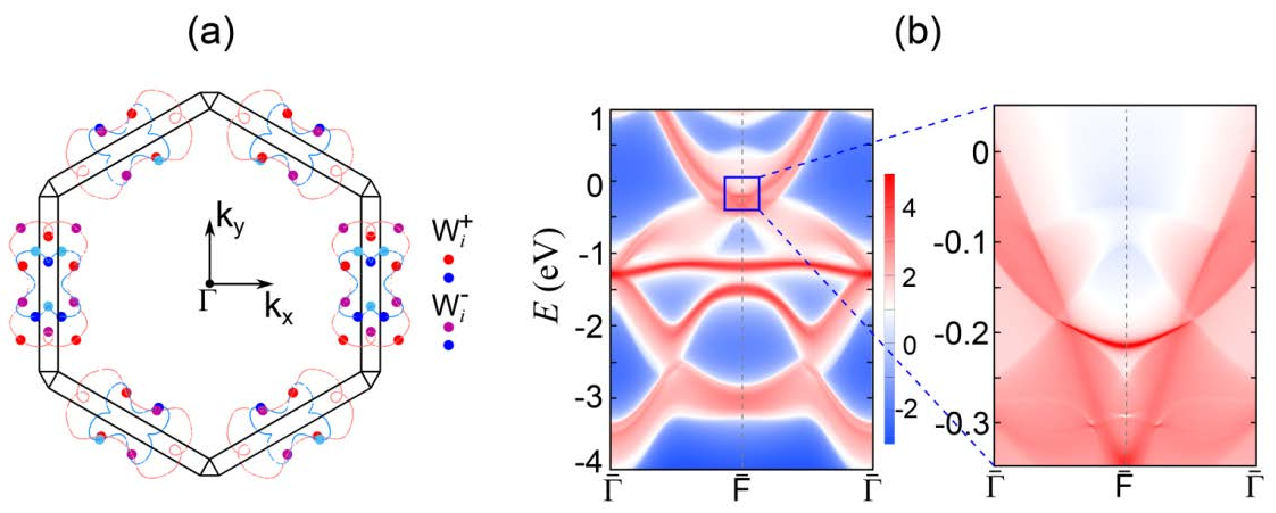}
\caption{(a) Distribution of Weyl nodes, obtained when the magnetic direction is along the $y$ axis. The projected surface spectrum for the (111) surface of Y$_2$C with the close-up image around the $\overline{F}$ point is given in (b). Here, the magnetic space group is reduced to C$_{2h}$ containing inversion symmetry $I$ and two-fold rotation symmetry C$_{2y}$. Although the positions of Weyls nodes change depending on the magnetization direction, the drumhead surface state around the $\overline{F}$ point remains intact.}
\end{figure}

\newpage
{\bf Table SI. Positions and energies of Weyl nodes in the first Brillouin zone.}\\
\begin{table}[ht]
\begin{tabular}{ccccc}
 Weyl node               & $k_x$           & $k_y$         & $k_z$           & Energy (eV) \\ \hline
 W$_{1}^{+}$(W$_{1}^{-}$)& 0.888 (-0.888)  &  0 (0)        & 0.404 (-0.404)  & -0.183 \\
 W$_{2}^{+}$(W$_{2}^{-}$)& -0.932 (0.932)  &  0 (0)        & -0.505 (0.505)  & -0.424 \\
 W$_{3}^{+}$(W$_{3}^{-}$)& -0.870 (0.870)  &-0.076 (0.076) & 0.404 (-0.404)  & -0.396 \\
 W$_{4}^{+}$(W$_{4}^{-}$)& -0.870 (0.870)  &0.076 -(0.076) & 0.404 (-0.404)  & -0.396 \\
 W$_{5}^{+}$(W$_{5}^{-}$)& 0.925 (-0.925)  &-0.206 (0.206) &-0.171 (0.171)   & -0.172 \\
 W$_{6}^{+}$(W$_{6}^{-}$)& 0.925 (-0.925)  &0.206 (-0.206) &-0.171 (0.171)   & -0.172 \\
 W$_{7}^{+}$(W$_{7}^{-}$)& -1.022 (1.022)  &-0.368 (0.368) &0.138 (-0.138)   &  0.337 \\
 W$_{8}^{+}$(W$_{8}^{-}$)& -1.022 (1.022)  &0.368 (-0.368) &0.138 (-0.138)   &  0.337 \\
\end{tabular}
\end{table}
\vspace{0.2cm}
Table SI. Positions and energies of Weyl nodes in the first Brillouin zone for Y$_2$C. The positions ($k_x$, $k_y$, $k_z$) are in unit of {\AA}$^{-1}$. Energies are relative to the Fermi energy $E_{\rm F}$.

\vspace{0.4cm}
{\bf Table SII. Spin and orbital magnetic moments of bulk Y$_2$C.}\\
\begin{table}[hb]
\begin{tabular}{cc}
    (a) & (b)\\
    \begin{tabular}{cccc}
    Ion   &   $m_x$   & $m_y$   & $m_z$   \\ \hline
    Y     &     0     &   0     & 0.108   \\
    Y     &     0     &   0     & 0.108   \\
    C     &     0     &   0     & -0.005  \\
    X     &     0     &   0     & 0.169   \\
    \end{tabular}
    &
    \begin{tabular}{cccc}
    Ion   &   $m_x$   & $m_y$   & $m_z$   \\ \hline
    Y     &     0     &   0     & -0.002   \\
    Y     &     0     &   0     & -0.002  \\
    C     &     0     &   0     & 0  \\
    X     &     0     &   0     & 0   \\
    \end{tabular}
\end{tabular}
\end{table}
\vspace{0.2cm}
Table SII. Calculated (a) spin and (b) orbital magnetic moments of bulk Y$_2$C with the inclusion of SOC. Here, the magnetization direction is along the $z$ axis. The magnitudes of orbital magnetic moments are found to be two orders smaller than those of spin magnetic moments.

\end{flushleft}

\end{document}